\documentclass[amstex,12pt,a4paper]{article}
\usepackage{amssymb}
\usepackage{amsmath}
\usepackage{graphicx}
\usepackage{calc}
\usepackage{bm}

\usepackage{dsfont}
\usepackage{authblk}
\usepackage{fancyhdr}
\begin{document}

\title{A Semi-Analytic Approach To Valuing Auto-Callable Accrual Notes}

\author[1,3]{V.~G.~Filev\thanks{veselin.filev@cloudrisk.co.uk; vfilev@stp.dias.ie}}
\author[2]{P.~Neykov\thanks{plamen.neykov@cloudrisk.co.uk}}
\author[1,4]{G.~S.~Vasilev\thanks{genko.vasilev@cloudrisk.co.uk; gvasilev@phys.uni-sofia.bg}}


\affil[1]{\small \it R\&D, CloudRisk Ltd \\
Narodno Subranie 9, 1000 Sofia, Bulgaria}
\affil[2]{\small \it HQ, CloudRisk Ltd \\
308 Spice Quay Heights, 32 Shad Thames, London, SE1~2YL, United Kindom}
\affil[3]{\small \it School of Theoretical Physics\\
Dublin Institute for Advanced Studies,
10~Burlington Road, Dublin 4, Ireland}

\affil[4]{\small \it Department of Physics, Sofia University, James~Bourchier~5~blvd, 1164~Sofia,~Bulgaria}

\date{}
\maketitle

\begin{abstract}
We develop a semi-analytic approach to the valuation of auto-callable structures with accrual features subject to barrier conditions. Our approach is based on recent studies of multi-assed binaries, present in the literature. We extend these studies to the case of time-dependent parameters. We compare numerically the semi-analytic approach and the day to day Monte Carlo approach and conclude that the semi-analytic approach is more advantageous for high precision valuation.
\end{abstract}

\newpage
\tableofcontents





\section{Introduction}

\label{sec:intro}

Auto-callable structures are quite popular in the world of \
structured products. On top of the auto-callable structure it is
common to add features related to interest payments. Hence,
combining range accrual instruments and auto-call options not only
leads to interesting conditional dynamics, but gives an
illustrative example of a typical structured product
ref.~\cite{Bouzoubaa}. In addition to the strong path dependence
of the coupons the instrument's final redemption becomes path
dependent too. Intriguingly, within the Black--Scholes world one
can obtain a closed form expression for the payoff of such a
derivative. On the other side one can also rely on a
straightforward Monte Carlo (MC) approach ref.~\cite{Glasserman}.
Often the interest payment features embedded in the instrument
accrue a fixed amount daily, related to some trigger levels of the
underlyings. The standard approach for valuation of such
instruments is a daily MC simulation. The goal of this paper is to
propose an alternative semi-analytic approach (SA), which in some
cases performs significantly better than the brute force day to
day MC evaluation ref.~\cite{Korn}. As we are going to show, the
complexity of the evaluation of the auto-call probabilities grows
linearly with the number of observation times of the instrument
and one may expect that at some point the MC approach would become
more efficient. However, even in these higher dimensional cases
the semi-analytic approach provides a better control of the
sensitivities of the instrument, since contrary to the MC approach
it does not rely on a numerical differentiation. A relevant question
is what are the pros and cons of the above methods - i.e SA vs MC.
We address this question performing a thorough numerical
investigation.

Technically our work is heavily based on ref.~\cite{SKIPPER}, where a valuation formula for multi-asset,
multi-period binaries is provided. In addition to applying theses studies to auto-callable
and range accrual structures, we extend the main result of ref.~\cite{SKIPPER} to the case of
time-dependent parameters: volatilities, interest rates and dividend yields.\footnote{To the best
of our knowledge, a closed formula for time-dependant parameters have not been presented in the literature.}

The paper is structured as follows: In section two we begin with a brief description of the type of derivative
instrument that we are studying.

In section three we develop the quasi-analytic approach, extending
the results of ref. \cite{SKIPPER} to the case of time dependent
deterministic parameters obtaining an expression for the
probability of an early redemption in terms of the multivariate
cumulative normal distribution. Building on this approach we
obtain similar expression for the payoff at maturity, subject to
elaborate conditions. In addition we calculate the payoff of the
coupons as a sum over multivariate barrier options ref.~\cite{Hull},
using the developed SA approach to represent the pay-off of the
latter in terms of multivariate cumulative normal distribution
ref.~\cite{Zhang}.

Finally, in section four we apply our approach to concrete examples. We implement numerically both the SA and MC approaches
and demonstrate the advantage of applying the SA approach to lower dimensional systems, especially when a high precision valuation is required.

\section{The instrument}

\label{sec:inst}

In this paper we analyse a type of instrument which combines the features of
range accrual coupons with auto-call options.

--The instrument is linked to the performance of two correlated assets $S_1$
and $S_2$.

--The instrument has a finite number $M$ of observation times $%
T_1,T_2,\dots,T_M$. If at the observation time $T_k$ both assets $S_i$ are
simultaneously above certain barriers $b_{i,\,k}$ the instrument redeems at
100\%. This is the auto-call condition. To shift the valuation time at zero
we define $\tau = T-t$ and discuss the observation times $%
\tau_1\,,\tau_2\,,\dots,\tau_M$.

--At the observation times the instrument pays coupons proportional to the
number of days, in the period between the previous observation time and the
present,\footnote{%
Or the valuation day for the first observation time.} in which both assets $%
S_i$ were above certain barriers $c_i$.

--If the instrument reaches maturity, it redeems at 100\% if both assets $S_i
$ are above certain percentage $\kappa$ of their spot prices at issue time $%
\bar S_i$. If at least one of the assets is bellow $\kappa\,\bar S_i$ the
instrument pays only a part proportional to the minimum of the ratios $%
S_i/\bar S_i$.

\section{Semi-analytic approach}

\label{sec:anal}

In this section we outline our semi-analytic approach. We begin by providing
a formula for the auto-call probability.

\subsection{Indicator functions and common notations}

\label{sec:indic}

Without loss of generality, it is assumed that the auto-callable structure
has two underlyings. On the set of dates are imposed trigger conditions
related to the auto-call feature. If the auto-call triggers have never been
breached at the observation dates the auto-callable structure matures at its
final maturity date. On the opposite case, if one of the auto-call triggers
have been breached the instrument auto-calls at this particular date and has
its maturity.

Let us denote with $P_{k}$ the probability to auto-call at observation time $%
\tau _{k}$. Note that this implies that at previous observation times the
spot prices of the two assets where never simultaneously above the barriers $%
b_{i}$. We introduce the following notations: $X_{i,\,k}$ labels the spot
value of the assets $S_{i}$ at observation time $\tau _{k}$.

Using the standard notations, if probability space $\left( \Omega ,\digamma ,%
\mathbf{P}\right) $ is given, and $A\in \digamma ,$ than the indicator
function is defined as $\mathbf{E}\left( \mathbf{1}_{A}\right) =P(A).$

Using the above definition, the auto-call probability at time $\tau _{k},$%
for the general case with $n$ underlying indices is then given by the
expectation related to some probability measure $Q$ of the indicator
function:

\begin{equation}\label{ind-1}
P_{k}=\mathbf{E}_{Q}\left( \mathbf{1}_{(X_{1,\,1}<\,b_{1,\,1})\cup
(X_{2,\,1}<\,b_{2,\,1}),\,(X_{1,\,2}<\,b_{1,\,2})\cup
(X_{2,\,2}<\,b_{2,\,2}),\,\dots ,\,(X_{1,\,k}<\,b_{1,\,k})\cap
(X_{2,\,k}<\,b_{2,\,k})}\right) \ .
\end{equation}%
In order to simplify the notation, hereafter we will omit the probability
measure $Q.$ For the case of two underlyings, we can also define also the
probability that the instrument will not auto-call after the first $k$
observation times:
\begin{equation}
\bar{P}_{k}=\mathbf{E}\left( \mathbf{1}_{(X_{1,\,1}<\,b_{1,\,1})\cup
(X_{2,\,1}<\,b_{2,\,1}),\,(X_{1,\,2}<\,b_{1,\,2})\cup
(X_{2,\,2}<\,b_{2,\,2}),\,\dots ,\,(X_{1,\,k}<\,b_{1,\,k})\cup
(X_{2,\,k}<\,b_{2,\,k})}\right) \ .
\end{equation}%
Note that at each observation time we have more than one possibilities
reflected in the $\cup $ operation.

For example the event $(X_{1,\,1}<\,b_{1})\cup (X_{2,\,1}<\,b_{2})$ can be
split into the three scenarios $(X_{1,\,1}<\,b_{1})\cap (X_{2,\,1}<\,b_{2})$%
, $(X_{1,\,1}<\,b_{1})\cap (X_{2,\,1}>\,b_{2})$, $(X_{1,\,1}>\,b_{1})\cap
(X_{2,\,1}<\,b_{2})$.

We could do a bit better if we define $\tilde{X}_{1,\,s}=X_{1,\,s}/b_{1,\,s}$
and $\tilde{X}_{2,\,s}=X_{2,\,s}/b_{2,\,s}$. Then the condition $%
(X_{1,\,1}<\,b_{1,\,1})\cup (X_{2,\,1}<\,b_{2,\,1})$ can be split into the
two conditions $(\tilde{X}_{1,\,1}<1)\cap (\tilde{X}_{1,\,1}\,{\tilde{X}%
_{2,\,1}}^{-1}<1)$, $(\tilde{X}_{2,\,1}\,{\tilde{X}_{1,\,1}}^{-1}<1)\cap (%
\tilde{X}_{2,\,1}<1)$. Therefore to evaluate $\bar{P}_{k}$ we need to sum
over $2^{k}$ possible scenarios, each scenario containing $2k$ conditions.
This requires summing over $2^{k}$ different $2k$-dimensional cumulative
multivariate normal distributions~\cite{SKIPPER}, which is computationally
overwhelming for large values of $k$. Fortunately, using de Morgan rules we
can substantially reduce the computational cost.

Let us denote by ${\cal E}_{i}$ the event $(X_{1,\,i}<\,b_{1})\cup
(X_{2,\,i}<\,b_{2})$, then the event $\bar{\cal E}_{i}$ is written as the single
scenario $(X_{1,\,i}>\,b_{1})\cap (X_{2,\,i}>\,b_{2})$.

Using the well known probability relation

\begin{eqnarray*}
P\left( \bigcap\nolimits_{i=1}^{n}{\cal E}_{i}\right)  &=&\sum\limits_{i}P\left(
{\cal E}_{i}\right) -\sum\limits_{i,j}P\left( {\cal E}_{i}\cup {\cal E}_{j}\right)
+\sum\limits_{i,j,k}P\left( {\cal E}_{i}\cup {\cal E}_{j}\cup {\cal E}_{k}\right) + \\
&&...+\left( -1\right) ^{n}P\left( \bigcup\nolimits_{i=1}^{n}{\cal E}_{i}\right)
\end{eqnarray*}

and DeMorgan's law

\[
\overline{\left( \bigcup\nolimits_{i=1}^{n}{\cal E}_{i}\right) }=\bigcap%
\nolimits_{i=1}^{n}\overline{{\cal E}_{i}}
\]

can be shown that

\begin{equation}\label{barPk}
\bar{P}_{k}=P\left( \bigcap_{s=1}^{k}{\cal E}_{s}\right)
=1+\sum_{s=1}^{k}\sum_{\sigma _{s}\in C_{s}^{k}}(-1)^{s}P\left(
\bigcap_{j=1}^{s}\bar{{\cal E}}_{\sigma _{s}(j)}\right) \ .
\end{equation}

where the second sum is over all (sorted in ascending order) combinations of
$k$ elements $s-$th class, $C_{s}^{k}$. Note that there are again $2^{k}$
different terms, however only the last term is $2k$-dimensional.\footnote{%
In general the number of $2s$-dimensional terms is $\binom{k}{s}$.}

In the same spirit we can obtain a formula for the auto-call probabilities $%
P_{k}$:
\begin{equation}
P_{k}=\sum_{s=0}^{k-1}\sum_{\sigma _{s}\in C_{s}^{k-1}}(-1)^{s}P\left(
\bigcap_{j=1}^{s}\bar{{\cal E}}_{\sigma _{s}(j)}\cap \bar{{\cal E}}_{k}\right) \ ,
\label{APk}
\end{equation}%
where we have used a convention: $\cap _{j=1}^{0}\bar{{\cal E}}_{\sigma
_{0}(j)}\cap \bar{{\cal E}}_{k}=\bar{{\cal E}}_{k}$. Equations (\ref{barPk}) and (\ref{APk}%
) can be rewritten in terms of indicator functions. For compactness it is
convenient to adopt the notations of ref.~\cite{SKIPPER}. We introduce a
multi-index notation denoting by $X_{I}$ the element $X_{i,\,s}$, where $%
I=1\,,\dots \,,n$ and $n$ is the number of all observed assets' prices. In
the case considered in equation (\ref{ind-1}) we have $n=2\,k$. Using
lexicographical order we can make the map explicit:
\begin{equation}
(i,s)\rightarrow I=I[i,\,s]=2\ast (s-1)+i\,  \label{ldxic}
\end{equation}%
where we have used that $i=1,2$. Next we define the following notation:
\begin{equation}\label{actionA}
(X^{A})_{j}={X}_{1}^{A_{j1}}\,\dots \,{X}_{n}^{A_{jn}}~~~~j=1\,,\dots \,,m\ ,
\end{equation}%
where $m$ is the number of barrier conditions and A is an $n\times m$
matrix. With these notations a general indicator function can be written as:
\begin{equation}
\mathbf{1}_{m}(S\,\mathbf{X}^{A}>S\,\mathbf{a})\,  \label{ind_func}
\end{equation}%
where $\mathbf{a}$ is a vector of barriers and to allow for different types
of inequalities we have introduced the $m\times m$ diagonal matrix $S$ whose
diagonal elements take the values $\pm 1$ ('$+$' for '$>$' and '$-$' for '$<$%
'). Equations (\ref{barPk}),(\ref{APk}) now become:
\begin{eqnarray}\label{indicator barP}
&&\bar{P}_{k}=\mathbf{E}\left( 1+\sum_{s=1}^{k}\sum_{\sigma _{s}\in
C_{s}^{k}}(-1)^{s}\mathbf{1}_{2s}(\mathbf{X}^{A(\sigma _{s})}>\mathbf{b}%
(\sigma _{s}))\right) \ ,  \\
\label{indicator P}
&&P_{k}=\mathbf{E}\left( \sum_{s=0}^{k-1}\sum_{\sigma _{s}\in
C_{s}^{k-1}}(-1)^{s}\mathbf{1}_{2s+2}(\mathbf{X}^{\tilde{A}(\sigma _{s})}>%
\mathbf{\tilde{b}}(\sigma _{s}))\right) \ ,
\end{eqnarray}%
where $\mathbf{A}(\sigma _{s})$, $\mathbf{b}(\sigma _{s})$, $\mathbf{\tilde{A%
}}(\sigma _{s})$, $\mathbf{\tilde{b}}(\sigma _{s})$, are $2k\times 2s$, $%
1\times 2s$, $2k\times 2(s+1)$, $1\times 2(s+1)$ matrices, respectively.
Their non-zero entries are:
\begin{eqnarray}
&&A(\sigma _{s})_{I[i,\,\sigma _{s}(j)],\,I[i,\,j]}=1,~~~(\mathbf{b}(\sigma
_{s}))_{I[i,j]}=b_{i,\,\sigma _{s}(j)}\ ,  \label{Asigma} \\
&&\tilde{A}(\sigma _{s})_{I[i,\,\sigma _{s}(j)],\,I[i,\,j]}=1,~~~(\mathbf{%
\tilde{b}}(\sigma _{s}))_{I[i,j]}=b_{i,\,\sigma _{s}(j)}\ , \\
&&~~~\text{for}~~~i=1,2~\text{and}~j=1,\dots ,s\ .  \nonumber \\
&&\tilde{A}(\sigma _{0})_{I[i,\,k],\,i}=1,~~~(\mathbf{\tilde{b}}(\sigma
_{0}))_{i}=b_{i,\,k}\ ,~~~\text{for}~~i=1,2\ .
\end{eqnarray}%
In equation (\ref{Asigma}) we have used the map (\ref{ldxic}). Note that it
is crucial that the combinations $\sigma _{s}$ are sorted in ascending order.

\subsection{A time-dependant valuation formula}

If we restrict ourselves to time independent deterministic parameters
(interest rate, dividend yield, volatility) we can directly apply the
formula derived in ref. \cite{SKIPPER} to calculate the indicator functions
in equations (\ref{indicator barP}) and (\ref{indicator P}). However, this is a
very crude approximation when dealing with long instruments this is why we
extend the results of ref.~\cite{SKIPPER} to the time dependent case. The
starting point is to model the dynamics of the asset $S_{i}$ with a
geometric Brownian motion:
\begin{equation}
\frac{dS_{i}}{S_{i}}=\left( r(s)-q_{i}(s)\right) ds+\sigma
_{i}(s)\,dW_{i}(s)\ ,  \label{dyn-1}
\end{equation}%
where $W_{i}$ are correlated Brownian motions with correlation coefficient $%
\rho _{ij}$. Indeed the integrated form of equation (\ref{dyn-1}) is:
\begin{equation}
S_{i}(\tau )=S_{i}^{(0)}\,\exp \left\{ \int\limits_{0}^{\tau }\left(
r(s)-q_{i}(s)-\frac{1}{2}\sigma _{i}(s)^{2}\right) ds+\int\limits_{0}^{\tau
}\sigma _{i}\,dW_{i}(s)\right\}
\end{equation}%
For the asset $i$ at time $T_{k}$ we can write:
\begin{equation}
\log \tilde{X}_{i,k}=\log x_{i}+\left( \bar{r}_{i,k}-\bar{q}_{i,k}-\frac{1}{2%
}\,\bar{\sigma}_{i,k}^{2}\right) \tau _{k}+\bar{\sigma}_{i,k}\,\sqrt{\tau
_{k}}\,Z_{i,k}\ ,  \label{Xik}
\end{equation}%
where $Z_{i,k}$ is given by:
\begin{equation}
Z_{i,k}=\frac{1}{\bar{\sigma}_{i,k}\sqrt{\tau _{k}}}\int\limits_{0}^{\tau
_{k}}\,\sigma _{i}(s)\,dW_{i}(s)\   \label{Zik}
\end{equation}%
and
\begin{eqnarray}
\bar{r}_{i,k} &=&\frac{1}{\tau _{k}}\int\limits_{0}^{\tau _{k}}ds\,r_{i}(s)\
,  \nonumber  \label{aver} \\
\bar{q}_{i,k} &=&\frac{1}{\tau _{k}}\int\limits_{0}^{\tau _{k}}ds\,q_{i}(s)\
, \\
{\bar{\sigma}}_{i,k}^{2} &=&\frac{1}{\tau _{k}}\int\limits_{0}^{\tau
_{k}}ds\,\sigma _{i}(s)^{2}\ .  \nonumber
\end{eqnarray}%
Following ref.~\cite{SKIPPER} we define the quantities:
\begin{eqnarray}
\mu  &=&\left( \bar{r}_{i,k}-\bar{q}_{i,k}-\frac{1}{2}\,\bar{\sigma}%
_{i,k}^{2}\right) \tau _{k}\ ,  \nonumber  \label{mu and Sigma} \\
\Sigma  &=&\text{diag\thinspace }(\bar{\sigma}_{i,k}\,\sqrt{\tau _{k}})\ .
\end{eqnarray}%
which are straightforward generalisations of the corresponding definitions
in the time independent case~\cite{SKIPPER}. A bit more involved is the
expression for the correlation matrix $R$ defined as:
\begin{equation}
R_{(i,k)(j,l)}\equiv \langle Z_{i,k}\,,Z_{j,l}\rangle \ .
\end{equation}%
Using equation (\ref{Zik}) and the formula:
\begin{equation}
\left\langle \int\limits_{0}^{\tau _{1}}\sigma
_{i}(s)\,dW_{i}(s)\,,\,\int\limits_{0}^{\tau _{2}}\sigma
_{j}(r)\,dW_{j}(r)\right\rangle =\rho _{ij}\int\limits_{0}^{\min (\tau
_{1},\tau _{2})}\sigma _{i}(\tau )\,\sigma _{j}(\tau )\,d\tau \ ,
\end{equation}%
we obtain:
\begin{equation}
R_{(i,k)(j,l)}=\frac{\rho _{ij}}{\sqrt{\tau _{k}\tau _{l}}\,\bar{\sigma}%
_{i,k}\bar{\sigma}_{j,l}}\int\limits_{0}^{\min (\tau _{k},\tau _{l})}\sigma
_{i}(\tau )\,\sigma _{j}(\tau )\,d\tau \ .
\end{equation}%
Next following ref.~\cite{SKIPPER} we define:
\begin{eqnarray}
\Gamma  &=&\Sigma \,R\,\Sigma ^{\prime }\ ,  \label{Gamma,D,C,d} \\
D &=&\sqrt{\text{diag}\left( A\,\Gamma \,A^{\prime }\right) }\ ,  \nonumber
\\
C &=&D^{-1}\left( A\,\Gamma \,A^{\prime }\right) D^{-1}\ ,  \nonumber \\
\mathbf{d} &=&D^{-1}\left[ \log (\mathbf{x}^{A}/\mathbf{a})+A\,\mu \right] \
.  \nonumber
\end{eqnarray}%
Here it is used that $x_{i,k}=x_{i}$ for all $k=1,\dots ,M$. In therms of
these quantities the indicator function is given by the same expression as
in ref.~\cite{SKIPPER}, but the underlying variables are given in eq. (\ref%
{Gamma,D,C,d}) and due to the time-dependence thay are different from those
given in the work ref.~\cite{SKIPPER},
\begin{equation}
\mathbf{1}_{m}(S\,\tilde{\mathbf{X}}^{A(\omega )}>S\,\mathbf{a})=\mathcal{N}%
_{m}(S\,\mathbf{d(\omega )},\,S\,C(\omega )\,S)\,\ ,  \label{formula}
\end{equation}%
where $\mathcal{N}_{m}$ is the cumulative multivariate normal distribution
(centred around zero).

Note that equations (\ref{mu and Sigma})--(\ref{formula}) are valid for any $%
n\times m$ matrix $A$ and any positive barrier vector $\mathbf{a}$.

\subsection{Auto-call probability and final payoff}

Applying equation (\ref{formula}) to calculate the auto-call probability $P_k
$ we obtain:
\begin{equation}  \label{Pknew}
P_k =\sum_{s=0}^{k-1}\sum_{\sigma_s \in C^{k-1}_s}(-1)^s\mathcal{N}_{2s+2}(%
\mathbf{d}(\sigma_s)\, ,C(\sigma_s))\ ,~~~k=1\,\dots M-1\ ,
\end{equation}
where $\mathbf{d}(\sigma_s)$ and $C(\sigma_s)$ are obtained by substituting $%
\tilde A(\sigma_s)$ and $\mathbf{b(\sigma_s)}$ from equation (\ref{Asigma})
into equation (\ref{Gamma,D,C,d}). Note that the index $k$ in equation (\ref%
{Pknew}) runs from one to $M-1$. The reason is that the last observation
time is the maturity.

Let us denote by $P_{\mathrm{mat}}$ the probability to reach maturity%
\footnote{%
Note also that $P_{\mathrm{mat}} = \bar P_{M-1}$}. Clearly we have:
\begin{equation}
P_{\mathrm{mat}}=1-\sum_{k=1}^{M-1}P_k\ ,
\end{equation}
The probability $P_{\mathrm{mat}}$ can be split into two contributions:
\begin{equation}  \label{Pmat}
P_{\mathrm{mat}}=P_{\mathrm{up}} + P_{\mathrm{down}}\,
\end{equation}
Where $P_{\mathrm{up}}$ is the probability to reach maturity with both
assets simultaneously above the barrier $\kappa\,\bar S_i$, and $P_{\mathrm{%
down}}$ is the probability at least one fo the assets to be bellow the
barrier. In fact the probability $P_{\mathrm{up}}$ is exactly $P_M$, hence
we can write:
\begin{equation}
P_{\mathrm{up}}=\sum_{s=0}^{M-1}\sum_{\sigma_s \in C^{k-1}_s}(-1)^s\mathcal{N%
}_{2s+2}(\mathbf{d}(\sigma_s)\, ,C(\sigma_s))\ .
\end{equation}
Clearly this also determines $P_{\mathrm{down}}$ as $P_{\mathrm{down}}=P_{%
\mathrm{mat}}-P_{\mathrm{up}}$. To calculate the payoff at maturity we also
need the average performance of the assets subject to the condition that the
worst performing asset is bellow the barrier $\kappa\,\bar{S_i}$. The
probability for this to happen is exactly $P_{\mathrm{down}}$, which is a
function of the parameter $\kappa$.

Let us denote $\hat X_i =S_i/\bar S_i$ and define $\hat X = \mathrm{min}%
(\hat X_1,\hat X_2)$, the probability $P_{\mathrm{down}}$ can be written as:
\begin{equation}
P_{\mathrm{down}} = P(\hat X < \kappa)\ .
\end{equation}
The average performance of the assets provided that at least one of the
assets is bellow the barrier $\kappa$ is then proportional to the
conditional expectation value $\langle \hat X\rangle |_{\hat X < \kappa}$:
\begin{equation}
\left\langle \mathrm{min}\left(\frac {S_1} {\bar S_1},\,\frac {S_2 }{\bar S_2%
}\right)\right\rangle \Big |_{\hat X < \kappa} =-\frac{1}{P_{\mathrm{down}}}%
\int\limits_0^\kappa d\kappa\, \kappa\,\frac {d P_{\mathrm{up}}}{d \kappa}\ ,
\end{equation}
where we have used that $dP_{\mathrm{mat}}/d\kappa = 0$. Therefore, the
payoff at maturity is given by:
\begin{equation}  \label{Vmat}
V_{\mathrm{maturty}}=P_{\mathrm{up}} +P_{\mathrm{down}}\left\langle \mathrm{%
min}\left(\frac {S_1} {\bar S_1},\,\frac {S_2 }{\bar S_2}\right)\right%
\rangle \Big |_{\hat X < \kappa} =P_{\mathrm{up}} - \int\limits_0^\kappa
d\kappa\, \kappa\,\frac {d P_{\mathrm{up}}}{d \kappa}\ ,
\end{equation}
In the next subsection we calculate the contribution of the coupons.

\subsection{Coupon contribution}
\label{section:coupons}
To obtain the total payoff we have to evaluate the contribution of the
coupons. This can be done by summing over a type of two-asset binary (cash-or-nothing)
options, conditional on the survival of the instrument to the appropriate accrual period. Indeed the probability at time $\tau $ both assets to be above the
barrier is given by the probability for such an option to pay. In the case of the first accrual period this reduced to the standard two-asset binary option~\cite{Heynen}.
To write down a closed form expression for this probability we need to add
one more observation time $\tau _{a}$, which will iterate over the accrual
dates. Clearly the simplest case is when $0\leq t_{a}\leq \tau _{1}$, that
is the first accrual period. In this case we apply formula (\ref{formula}),
for just one observation time $\tau _{a}$, with $A=S=\mathbf{1}_{2}$ and $%
\mathbf{a}=\mathbf{c}$. In more details the probability the coupons to pay
at time $\tau _{a}<T_{1}$, $P_{01}(\tau _{a})$ is given by:
\begin{eqnarray}
P_{01}(\tau _{a}) &=&\mathcal{N}_{2}(\mathbf{d}_{2}(\tau _{a}),C_{2})\ ,
\nonumber  \label{P01} \\
d_{i} &=&\frac{\log (\bar{S}_{i}/c_{i})+(\bar{r}-\bar{q}_{i}(\tau _{a})-\bar{%
\sigma}_{i}(\tau _{a})^{2}/2)\,\tau _{a}}{{\bar{\sigma}_{i}}(\tau _{a})\sqrt{%
\tau _{a}}}\ , ~~~i=1,2\ ,\\
C_{2} &=&\left(
\begin{array}{cc}
1 & \rho _{12} \\
\rho _{21} & 1%
\end{array}%
\right)
\end{eqnarray}%
where $\bar{r},\bar{q}_{i}(\tau _{a})$ and $\bar{\sigma _{i}}(\tau _{a})$
are given by equations (\ref{aver}) with $\tau _{k}=\tau _{a}$. The total
number of days in which coupons have been payed in the period $0$ to $\tau
_{1}$, $N_{1}$ is then given by:
\begin{equation}
N_{1}=\sum_{\tau _{a}\,=\,1}^{\tau _{\,1}}P_{01}(\tau _{a})\ .  \label{N1}
\end{equation}%
In the same way we can obtain a formula for the number of coupon days in the
second accrual period. The only difference is that now in addition to the
condition both assets to be above the accrual barrier we also have the
condition that the instrument did not auto-call at time $\tau _{1}$. In
general the probability the coupons to pay at time $\tau _{a}$ in the $k$-th
accrual period is the joint probability that the instrument did not
auto-call at the first $k-1$ observation times and both assets are above the
accrual barrier at time $\tau _{a}$. Denoting by ${\cal E}^C_{\tau _{a}}$ the event
that the assets are above the accrual barrier at time $\tau _{a}$ and using
the notations from section \ref{sec:indic}, one can show that\footnote{%
The derivation is analogous to that of equation (\ref{APk}).}:
\begin{equation}
P_{k-1,k}(\tau _{a})=\sum_{s=0}^{k-1}\sum_{\sigma _{s}\in
C_{s}^{k-1}}(-1)^{s}P\left( \bigcap_{j=1}^{s}\bar{{\cal E}}_{\sigma _{s}(j)}\cap
{\cal E}^C_{\tau _{a}}\right) \ ,  \label{APk-1k}
\end{equation}%
where again we have used the convention: $\cap _{j=1}^{0}\bar{{\cal E}}_{\sigma
_{0}(j)}\cap {\cal E}^C_{\tau _{a}}= {\cal E}^C_{\tau _{a}}$. Equation (\ref{APk-1k}) can be
rewritten in analogy to equation (\ref{indicator P}) as:
\begin{equation}
P_{k-1,k}(\tau _{a})=\mathbf{E}\left( \sum_{s=0}^{k-1}\sum_{\sigma _{s}\in
C_{s}^{k-1}}(-1)^{s}\mathbf{1}_{2s+2}(\mathbf{\tilde{X}_{\sigma _{s}}}>%
\mathbf{bc_{\sigma _{s}}})\ \right) ,
\end{equation}%
where $\mathbf{\tilde{X}_{\sigma _{s}}}$ is the vector: $[X_{1,\sigma
_{s}(1)},\,X_{2,\sigma _{s}(1)},\,\dots \,,X_{1,\sigma
_{s}(s)},\,X_{2,\sigma _{s}(s)},\,S_{1}(\tau _{a}),\,S_{2}(\tau _{a})]$ and $%
\mathbf{bc_{\sigma _{s}}}$ is the vector: $[b_{1,\sigma
_{s}(1)},\,b_{2,\sigma _{s}(1)},\,\dots \,,b_{1,\sigma
_{s}(s)},\,b_{2,\sigma _{s}(s)},\,c_{1},\,c_{2}]$. Denoting by $\tilde{C}%
_{\sigma _{s}}(\tau _{a})$ the covariant matrix constructed using equations (%
\ref{aver})-(\ref{Gamma,D,C,d}) with times $\tau _{\sigma _{s}(1)},\,\dots
,\,\tau _{\sigma _{s}(s)},\,\tau _{a}$ and denoting by $\mathbf{\tilde{d}%
_{\sigma _{s}}}$ the corresponding quantity in equation (\ref{Gamma,D,C,d})
constructed using the barrier vector $\mathbf{bc_{\sigma _{s}}}$, we can
write:
\begin{equation}
P_{k-1,k}(\tau _{a})=\sum_{s=0}^{k-1}\sum_{\sigma _{s}\in
C_{s}^{k-1}}(-1)^{s}\mathcal{N}_{2s+2}(\mathbf{\tilde{d}_{\sigma _{s}}}\,,%
\tilde{C}_{\sigma _{s}}(\tau _{a}))\ .
\end{equation}%
For the number of coupon paying days in the $k$-th accrual period we obtain:
\begin{equation}
N_{k}=\sum_{\tau _{a}\,=\,\tau _{k-1}\,+\,1}^{\tau _{k}}\,P_{k-1,k}(\tau
_{a})\ .
\end{equation}%
To calculate the contribution of the coupons to the total payoff we need to
take into account the discount factors, since we have assumed that the
coupons are payed at the observation times\footnote{%
Note that in practise there are a separate payment dates shortly after the
corresponding observation date.}. Note that the probability the coupons to pay
already include the probability to reach that accrual period. Therefore, the total
coupon contribution is given by:

\begin{equation}
VC_{M}=\gamma \,\sum_{s=1}^{M}\,e^{-\bar{r}_{s}\,\tau _{s}}\,N_{s}\ ,
\end{equation}%
where $\gamma $ is the daily rate of the coupon.

\subsection{Total payoff}

Assuming for simplicity that the instrument redeems at 100 \% in the event
of an auto-call (which in reality is quite common), for the total payoff we
obtain:
\begin{eqnarray}
V_{\mathrm{tot}}&=&V_{\mathrm{maturity}}+\sum_{k=1}^{M-1}e^{-\bar{r}_{k}\,\tau _{k}}\,P_k+VC_M
\end{eqnarray}
where we have substituted $P_{\mathrm{mat}}$ from equation (\ref{Pmat}).

\section{Applications}

In this section we outline some of the applications of the formalism
developed above. We begin with the simplest case of a pure accrual
instrument.

\subsection{Pure accrual instrument}

The pure accrual instrument that we consider in this subsection has the
following characteristics:

--It pays a daily coupon at rate $\gamma$ if at closing time both assets $S_i
$ are above the accrual barriers $c_i$

--At maturity (time $\tau_m$), it redeems at 100\% if both assets $S_i$ are
above certain percentage $\kappa$ of their spot prices at issue time $\bar
S_i$. If at least one of the assets is bellow $\kappa\,\bar S_i$ the
instrument pays only a part proportional to the minimum of the ratios $%
S_i/\bar S_i$.

Clearly this is the general instrument that we considered with the auto-call
option removed. In this simple case the semi-analytic approach of section~%
\ref{sec:anal} is particularly efficient. The coupons are calculated by the
first period formulas in equations (\ref{P01}), (\ref{N1}) with $\tau_a =
\tau_m$, while the payoff at maturity is calculated using equation (\ref%
{Vmat}), with $P_{\mathrm{up}}$ given by:
\begin{equation}
P_{\mathrm{up}} = \mathcal{N}_2(\mathbf{\tilde d}_2(\tau_m), C_2) \ ,
\end{equation}
where $C_2$ is given in equation (\ref{P01}) and $\mathbf{\tilde d}_2(\tau_m)
$ is given by:
\begin{equation}
d_{i} =\frac{\log (\bar{S}_{i}/c_{i})+(\bar{r}-\bar{q}_{i}(\tau _m)-\bar{%
\sigma}_{i}(\tau_m)^{2}/2)\,\tau_m}{{\bar{\sigma}_{i}}(\tau_m)\sqrt{%
\tau_m}}\ , ~~~i = 1,2\ ,
\end{equation}
where $\bar{r},\bar{q}_{i}(\tau_m)$ and $\bar{\sigma _{i}}(\tau_m)$
are given by equations (\ref{aver}) with $\tau _{k}=\tau_m$.

\subsection{Dual index range accrual autocallable instrument}
In this section we compare the efficiency of our semi-analytic (SA) approach and that of a standard Monte Carlo (MC) approach. Since the dimensionality of the SA problem increases linearly with the number of
the auto-call dates, we consider the case of one auto-call date and two range accrual periods. Therefore, our problem is four dimensional and we would still need to rely on numerical methods to estimate the cumulative distributions.

To simplify the analysis even further and facilitate the comparison, we simplify the pay-off at maturity. The instrument pays 100\% if both underlyings perform above the final barrier $\kappa$ (as before), but if this condition is not satisfied, instead of redeeming a worse performance: $\mathrm{min}\left({S_1} /{\bar S_1},\, {S_2 }/{\bar S_2}\right)$ fraction, the instrument redeems at $\kappa \times 100\, \%$. Equation (\ref{Vmat}) then simplifies to:
\begin{equation}
V_{\mathrm{maturty}}=P_{\mathrm{up}} +\kappa\,P_{\mathrm{down}}\ .
\end{equation}
The description of the coupon payments remains the same as in section \ref{section:coupons}. The volatilities $\sigma_i$, dividend yields $q_i$ , interest rate $r$ and correlation correlation coefficient $\rho$ used in the numerical example are presented in table 1. In addition the final barrier was set at 60\% ($\kappa = 0.60$) and the daily accrual rate used was (15/365)\% ($\gamma = 0.15/365$). The length of each accrual period was one year so that: $\tau_1 =1$ and $\tau_2 =2$.
\begin{table}[h]
\begin{center}
\begin{tabular}{c|c|c|c}
$\sigma_i$ & $q_i$ & $r$ & $\rho$ \\
\hline
0.25 & 0.005 & 0.01 & 0.78 \\
0.20 & 0.007 & 0.01 & 0.78
\end{tabular}
\end{center}
\caption{\small Volatilities $\sigma_i$, dividend yields $q_i$, interest rate $r$ and correlation $\rho$ used in the numerical example. }
\end{table}%

\begin{figure}[h] 
   \centering
   \includegraphics[width=5.5in]{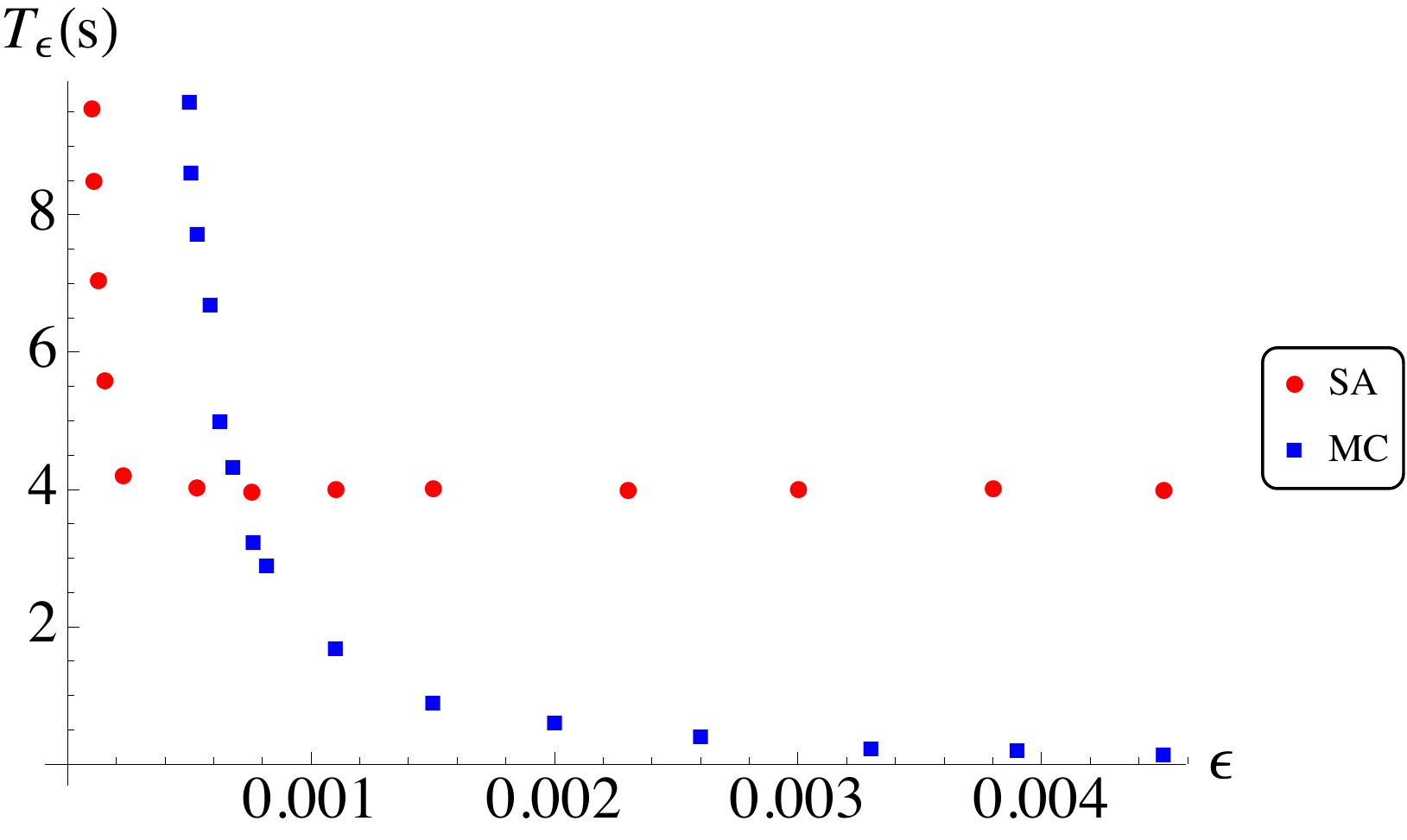}
 \caption{\small A plot of the running time $T_\epsilon$ in seconds as a function of the absolute error $\epsilon$. The round red dost represent the SA results, while the square blue dost correspond to the MC data.}
   \label{fig:1}
\end{figure}

To compare the efficiency of the algorithms we compared the running times $T_{\epsilon}$ as functions of the absolute error $\epsilon$. The resulting plot is presented in figure \ref{fig:1}.
The round dots correspond to the SA approach, while the square points represent the MC data. As one may expect, the running time $T_\epsilon$ for the MC algorithm increases as $\sim1/\epsilon^2$ and while negligible for $\epsilon < 0.01$, it increases rapidly to $\sim 10 s$, for $\epsilon = 5.0 \times 10^{-4}$. On the other side the SA method has a steady computation time $T_\epsilon \sim 4 s$, for $\epsilon < 2.0\times 10^{-4}$. The SA and MC curves intersect at $\epsilon \approx 0.7 \times 10^{-3}$. The advantage of using the SA method for higher precision  $\epsilon < 0.7 \times 10^{-3}$ is evident. For example  a calculation with $\epsilon = 2.0 \times 10^{-4}$ would require running the MC simulation for roughly $\sim 60\, s$, while the same accuracy can be achieved by the SA method for $\sim 5\, s$, which is a factor of twelve. Clearly the comparison depends on the implementation and the choice of parameters. To make the comparison fair we used MatLab for both methods. Using a vectorised MC algorithm for the Monte Carlo part and the built in MatLab cumulative distribution functions for the SA approach.

Another obvious advantage of the SA approach is the higher precision in the estimation of the sensitivities of the instrument. Semi-analytic expressions could be derided for most of the greeks, which enables their calculation with a limited numerical effort. This is clearly not the case in the MC approach, where one usually relies on a numerical differentiation.

Finally, as we pointed out at the beginning of this section the dimensionality of the problem increases linearly with the number of auto-call times. It is therefore expected that at some point the MC approach would become more efficient. Nevertheless, the SA approach could still be more efficient if the sensitivities are difficult to analyse in the MC approach.

\section{Conclusion}

This paper makes several contributions to the related literature.

Our main result is the development of a semi-analytical valuation method for
auto-callable instruments embedded with range accrual structures. Our approach
includes time-dependent parameters, and hence greatly facilitates practitioners.
In the process we extend the valuation formula for multi-asset, multi-period binaries
of ref.~\cite{SKIPPER} to the case of time-dependent parameters,
which the best of our knowledge is a novel result.

Another merit of this work is the comparison between the straightforward Monte Carlo
and the semi-analytical approaches.Our comparison shows that the semi-analytical
approach becomes more advantageous at higher precisions and is potentially order
of magnitude faster than the brute force Monte Carlo method. The semi-analytical
approach is also particularly useful when calculating the sensitivities of the instrument.
It is widely accepted that the sensitivity calculations are often more important than the
instrument price itself, due to their contribution for the correct instrument hedging.

Finally, our work can be used as a starting point for modelling
more complex structures related to range accrual auto-callable
instruments. Furthermore, although the numerical
examples and the presented formulas are given for the
two-dimension cases, multi-asset and multi-period generalisation
of the formulas can be easily written using the key formulas
presented here.

\paragraph{Acknowledgements:}

We would like to thank Bojidar Ibrishimov for critically reading
the manuscript.

\appendix

\section{Proof of the valuation formula}

For completeness we provide a proof of formula (\ref{formula}). Our proof follows the steps outlined in reference \cite{SKIPPER}. Using the definitions (\ref{actionA}),  (\ref{mu and Sigma}) and equation (\ref{Xik}) it is easy to obtain:
\begin{equation}
\log {\bf\tilde X}^A=\log{\bf x}^A +A  \,{\bf \mu} + A\,\Sigma\,{\bf Z}\ .
\end{equation}
Furthermore, the monotonicity of the logarithmic function implies:
\begin{equation}
\mathds{1}_m(S\, {\bf X}^{A} > S\,{\bf a}) = \mathds{1}_m(S\, \log{\bf X}^{A} > S\,\log{\bf a}) = \mathds{1}_m({\bf B \, Z} < {\bf b})\, \,
\end{equation}
where:
\begin{eqnarray}\label{B and b}
{\bf B} &=& -S\,A\,\Sigma\ , \\
{\bf b} &=& S\,(\log{\bf x}^A/{\bf a} +A\,{\bf \mu})\ .
\end{eqnarray}
Now we use a Lemma from ref.~\cite{SKIPPER} (which we will prove for completeness):
\\
{\bf Lemma 1.}
{\it
If {\bf B} is an $\bf m\times n$ matrix of rank ${\bf m}\leq\bf n$ and ${\bf Z}$ is a random unit variate vector of length $\bf n$ with correlation matrix $\bf R$. Then:
\begin{equation}\label{lemma}
E\left\{\mathds{1}_m(B{\bf\,Z} < {\bf b})\right\} = {\cal N}_m (D^{-1}{\bf b},\, D^{-1}(B\,R\,B^{T})\,D^{-1})\, \ ,
\end{equation}
where:
\begin{eqnarray}
D = \sqrt{{\rm diag}(B\,R\,B^T)}\ .
\end{eqnarray}
}

Applying Lemma 1 for ${\bf B}$ and $\bf b$ given in equation (\ref{B and b}), we obtain:
\begin{eqnarray}\nonumber\label{relations}
&&D = {\rm diag}(S\,A\,\Sigma\,R\,\Sigma^T\,A^T\,S) = {\rm diag}(A\,\Sigma\,R\,\Sigma\,A^T)={\rm diag}(A\,\Gamma\,A^T)\ ,\\
&&D^{-1}\,(S\,A\,\Sigma\,)R\,(\Sigma\,A^T\,S)\,D^{-1} = S\,D^{-1}\,A\,\Sigma\,R\,\Sigma\,A^T\,D^{-1}\,S =S\,C\,S \ , \nonumber\\
&&D^{-1}\,{\bf b}  =S\,D^{-1}\,(\log{\bf x}^A/{\bf a} +A.{\bf \mu}) =S\,{\bf d}\ ,
\end{eqnarray}
where we have used that $S$ and $D$ are diagonal and commute and that $S_{i,i}^2=1$. Substituting relations (\ref{relations}) into equation (\ref{lemma}) we arrive at equation (\ref{formula}). Now let us prove Lemma 1:\\
{\bf Proof:} Let us complete the $m\times n$ matrix $B$ to an $n\times n$ non-singular matrix $\tilde B$. We write:
\begin{equation}
\tilde B =    \begin{bmatrix} 
      B  \\
      B_{\perp}
   \end{bmatrix}\ ,
\end{equation}
where $B_{\perp}$ is an $(n-m)\times n$ matrix, which we are going to specify bellow. Consider the Cholesky decomposition of the correlation matrix $R$:
\begin{equation}
R = U\,U^T\ .
\end{equation}
Next we transform the matrix $\tilde B$ with $U$ via $\tilde B' = \tilde B\, U$, which implies:
\begin{eqnarray}
B' &=& B\, U\ , \\
B_{\perp}' &=& B_{\perp}\,U\ .
\end{eqnarray}
Since $B$ has rank $m$ and $U$ is invertible, $B'$ also has a rank $m$. We can therefore think of $B'$ as $m$ independent $n-column$ vectors. Spanning an $m$-dimensional subspace ${\cal L}^m$. We are always free to choose $B_{\perp}'$ to be a matrix of $n-m$ independent $n-column$ vectors spanning the orthogonal completion of ${\cal L}^m$. Making this choice of $B_{\perp}'$ implies:
\begin{equation}\label{off_zero}
B_{\perp} \,R \,B^T = B_{\perp}\,U\,(B\,U)^T = B_{\perp}' {B'}^T = 0\ ,
\end{equation}
Next we apply the transformation ${\bf Y} = \tilde B\, {\bf Z}$. The covariance matrix of the random vector $\bf Y$ is given by:
\begin{equation}
C = \tilde B \, R\,\tilde B^T =
\begin{bmatrix} 
      B\,R\,B^T & 0  \\
      0 & B_{\perp}\, R\, B_{\perp}^T
   \end{bmatrix}\ ,
\end{equation}
where we have used equation (\ref{off_zero}). Defining:
\begin{equation}
{\bf Y_{||}} = B\,{\bf Z} \ , ~~~~{\bf Y_{\perp}} = B_{\perp}\,{\bf Z} \ ,
\end{equation}
the condition $\mathds{1}_m(B{\bf\,Z} < {\bf b})$ becomes $\mathds{1}_m({\bf Y_{||}}< {\bf b})$. Furthermore, the probability density function of $\bf Y$ factorises:
\begin{eqnarray}
\rho({\bf Y}) &=& \frac{1}{(2\pi)^{n/2}\sqrt{\det (\tilde B\,R\,\tilde B^T)}}\exp\left(-\frac{1}{2}{\bf Y}^T\,(\tilde B\, R\tilde B^T)^{-1}\,{\bf Y}\right) =\nonumber \\
&=&\frac{1}{(2\pi)^{(n-m)/2}\sqrt{\det (B_{\perp}\,R\,B_{\perp}^T)}}\exp\left(-\frac{1}{2}{\bf Y}_{\perp}^T\,(B_{\perp}\, R\,B_{\perp}^T)^{-1}\,{\bf Y}_{\perp}\right)\times\nonumber\\
&\times&\frac{1}{(2\pi)^{m/2}\sqrt{\det (B\,R\,B^T)}}\exp\left(-\frac{1}{2}{\bf Y}_{||}^T\,(B\, R\,B^T)^{-1}\,{\bf Y}_{||}\right) = \nonumber \\
&=&\rho_{\perp}({\bf Y}_{\perp})\times \rho_{||}({\bf Y}_{||})
\end{eqnarray}
Since there are no conditions imposed on ${\bf Y}_{\perp}$ the integral over $\rho_{\perp}({\bf Y}_{\perp})$ is simply unity. What remains is the integral over $\rho_{||}({\bf Y}_{||})$, which upon the normalisation: ${\bf Y}_{\perp}\to D^{-1/2}{\bf Y}_{\perp}$ gives equation (\ref{lemma}).

\iffalse
\bibliographystyle{plain}
\bibliography{}

\begin{thebibliography}{9}

\bibitem{Bouzoubaa} M. Bouzoubaa and A. Osseiran (2010). "Exotic Options and Hybrids - A Guide to Structuring, Pricing and Trading",
\textit{2010 John Wiley \& Sons, Ltd},

\bibitem{Glasserman} P. Glasserman (2003). "Monte Carlo Methods in Financial Engineering",
\textit{2003 Springer },

\bibitem{Korn} R. Korn, E. Korn and G. Kroisandt (2010). "Monte Carlo Methods and Models in Finance and Insurance",
\textit{2010 Taylor and Francis Group, LLC},

\bibitem{SKIPPER} Max Skipper and Peter Buchen (2009). "A valuation formula
for multi-asset, multi-period binaries in a Black--Scholes
economy". The ANZIAM Journal, 50, pp 475-485.
doi:10.1017/S1446181109000285,

\bibitem{Hull} J. Hull (2015). "Options, Futures, and other Derivatives, 9ed.",
\textit{2015, 2012, 2009 Pearson Education},

\bibitem{Zhang} P. Zhang (1998). "Exotic Options, 2ed.",
\textit{1998 World Scientific},

\bibitem{Heynen} R. C. Heynen and H. M. Kat (1996). "Brick by Brick",
\textit{Risk Magazine},~9(6).


\end{thebibliography}
\else

\fi

\end{document}